# Absolute diffusion measurements of active enzyme solutions by NMR

Jan-Philipp Günther,[1,2] Günter Majer,[1,*] Peer Fischer[1,2,*]

[1]Max Planck Institute for Intelligent Systems, Heisenbergstr. 3, 70569 Stuttgart, Germany

[2]Institute of Physical Chemistry, University of Stuttgart, Pfaffenwaldring 55, 70569 Stuttgart, Germany

*Authors to whom correspondence should be addressed: Günter Majer (majer@is.mpg.de), Peer Fischer (fischer@is.mpg.de)

**Abstract**

The diffusion of enzymes is of fundamental importance for many biochemical processes. Enhanced or directed enzyme diffusion can alter the accessibility of substrates and the organization of enzymes within cells. Several studies based on fluorescence correlation spectroscopy (FCS) report enhanced diffusion of enzymes upon interaction with their substrate or inhibitor. In this context, major importance is given to the enzyme fructose-bisphosphate aldolase, for which enhanced diffusion has been reported even though the catalysed reaction is endothermic. Additionally, enhanced diffusion of tracer particles surrounding the active aldolase enzymes has been reported. These studies suggest that active enzymes can act as chemical motors that self-propel and give rise to enhanced diffusion. However, fluorescence studies of enzymes can, despite several advantages, suffer from artefacts. Here we show that the absolute diffusion coefficients of active enzyme solutions can be determined with Pulsed Field Gradient Nuclear Magnetic Resonance (PFG-NMR). The advantage of PFG-NMR is that the motion of the molecule of interest is directly observed in its native state without the need for any labelling. Further, PFG-NMR is model-free and thus yields absolute diffusion constants. Our PFG-NMR experiments of solutions containing active fructose bisphosphate aldolase from rabbit muscle do not show any diffusion enhancement for the active enzymes nor the surrounding molecules. Additionally, we do not observe any diffusion enhancement of aldolase in the presence of its inhibitor pyrophosphate.

# I. Introduction

Chemical reactions on the surface of microparticles can cause local concentration gradients around the particle. A local concentration gradient of substrate or product molecules can in turn give rise to fluid flows around the particle, and momentum conservation then requires that the particle moves in the opposite direction to these flows. Energy is consumed, which powers the self-propelled particles. Such "active" particles are also known as chemical motors with interesting properties and applications.[1,2,3,4,5,6,7] Whether these or related mechanisms can also influence the diffusive behavior of nanometer-sized catalytically-active materials is not yet fully understood.[8,9] Experimental diffusion studies of active molecular machines,[10] biopolymers,[11] and small nanocolloids[12] suspended in solution are challenging and lead to different results regarding the existence and magnitude of the diffusion enhancement in these nanosystems. The question whether the reaction of enzymes can cause their self-propulsion has potentially important implications for biochemistry. Recent studies suggest that the diffusion of active enzymes differs from their purely passive Brownian motion.[13,14,15,16,17,18,19] These studies rely on measurements of fluorescently labeled enzymes. However, it was recently shown that photo-physical artefacts can feign apparent diffusion enhancement of the enzyme when it is active.[11] It is therefore helpful to consider alternative experimental methods that do not require fluorophores and are thus independent of any photo-physical effects. In addition, low concentration of enzymes in fluorescence measurements can be problematic as it may cause enzyme oligomers to dissociate.

Nuclear magnetic resonance (NMR) offers unique possibilities to measure diffusion coefficients by combining pulsed field gradients (PFGs) with either a spin-echo or stimulated echo sequences. PFG-NMR has proven to be a powerful tool to measure absolute values of the diffusion coefficients of solutes directly, i.e. without the need for fluorescence labeling and independent of any diffusion-model assumptions.[20,21] Unlike fluorescence correlation spectroscopy (FCS), PFG-NMR does not need any calibration or reference measurement.[22] The diffusion constant of several molecules can be determined simultaneously, provided the molecules possess distinct NMR peaks, which is often the case.[20,21] This makes PFG-NMR a pseudo-2D-NMR method also often referred to as diffusion ordered NMR spectroscopy (DOSY).[23] We like to emphasize that in any diffusion measurement it is also possible to measure the diffusion of the solvent, which provides an internal reference and control of the accuracy of the diffusion measurements.[24] In particular, it is possible to determine the viscosity of any reaction mixture via the experimentally determined diffusivity of traces of protons (mainly HDO) in D$_2$O, using the Stokes-Einstein-Relation:

$$D = \frac{k_B T}{6\pi\eta R} \quad (1)$$

Here, we use PFG-NMR measurements to study the diffusive behavior of the enzyme fructose-bisphosphate aldolase (ALD) in its active state in various solutions. Enhanced diffusion of up to 35 %[18] has been reported from FCS measurements for ALD upon substrate conversion. It has also been reported that ALD can cause the enhanced diffusion of other molecules present in solution.[16] Additionally, it has been reported that ALD shows 20 % diffusion enhancement upon interaction with its inhibitor.[17] The theoretical explanations regarding the potential origin and magnitude of these effects differ.[25,26,27,28] PFG-NMR has been used to study the diffusion of small molecular tracers in the presence of passive proteins[24] as well as active molecular catalysts.[29] However, to the best of our knowledge, this is the first application of PFG-NMR to determine the diffusion of active enzymes.

PFG-NMR principle

Using the stimulated echo sequence depicted in Fig. 1, the nuclei of the molecule of interest are "spin-labeled" in the $z$-direction by a magnetic field gradient $G$ along $z$, which is applied for a time $\delta_S$. The signal $I$ acquired at the end of the stimulated echo sequence depends on whether the nuclei move away from their original position (along the $z$-direction). The maximum signal $I_0$ is measured in the absence of diffusion. If the molecule and hence its nuclei change position due to motion, the attenuation of the signal is expressed by the following equation[20,21]

$$I(G) = I_0 \cdot \exp(-b(G) \cdot D), \quad (2)$$

where $D$ denotes the diffusion coefficient of the molecule and $I(G)$ is the integrated intensity of a resonance peak in the NMR spectrum. $b(G)$ is an experimental parameter that depends on the diffusion time $\Delta$ as well as the length and amplitude of the gradient pulse. If the shape of the gradient pulse is given by a half-sine function with a duration $\delta_S$, then $\delta_G = \frac{2}{\pi}\delta_S$ where $\delta_G$ is the effective gradient time. In this case, the value $b(G)$ in Eq. (1) is given by[30,31,32]

$$b(G) = \gamma^2 G^2 \left[\delta_G^2 \left(\Delta - \frac{\pi}{8}\delta_G\right)\right]. \quad (3)$$

The diffusion time $\Delta$ is given by the separation of the leading edges of the gradient pulses und $\gamma = 2\pi \cdot 42.576$ MHz/T denotes the gyromagnetic ratio of the protons for $^1$H-PFG-NMR.

In this work, all parameters except $G$ were kept constant throughout each PFG-NMR experiment. By varying $G$ the attenuation of $I(G)$ can be fitted with Eq. 1 to reveal the diffusion coefficient of the molecule (see Fig. 1 (b)). Correspondingly, a linear regression of $\ln(I(G))$ as a function of $b(G)$ reveals the diffusion coefficient as a slope.[20,21]

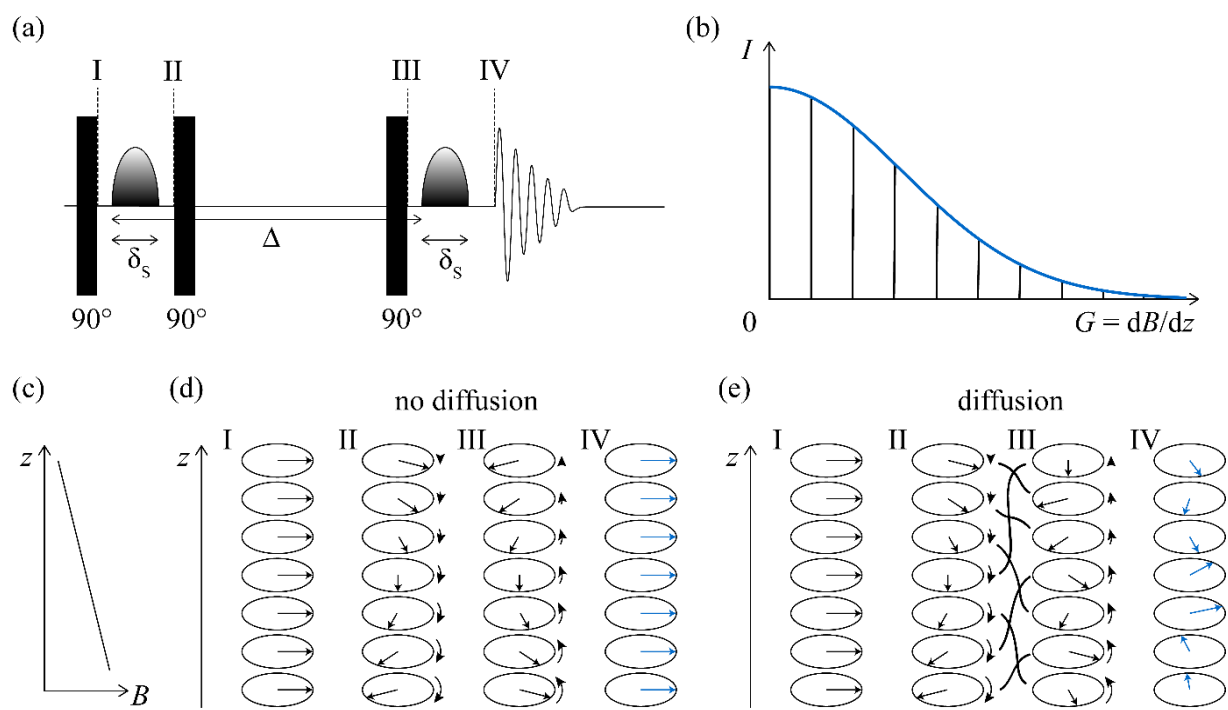

FIG. 1. General principle of Pulsed Field Gradient NMR experiments. (a) Stimulated spin echo pulse sequence including diffusion time $\Delta$ and half-sine gradient length $\delta_S$. (b) Dependence of the spin echo attenuation on the gradient strength $G$. (c) The gradient of the magnetic field $B$ is applied along the $z$-axis of the sample. (d) Spin evolution of nuclei at different $z$ positions during the NMR sequence. (e) Spin evolution during the NMR experiment including the effect of diffusion.

Proteins exhibit a wide distribution of chemical shifts in $^1$H-NMR spectroscopy (Fig. 2 (a)).[24] $^{13}$C- or $^{15}$N-NMR spectroscopy of proteins yields more distinct NMR spectra, but $^1$H nuclei exhibit the highest gyromagnetic ratio and abundance in proteins. Therefore, we decided to use the $^1$H signal together with wide signal integration for our protein diffusion studies. Accordingly, all samples have been prepared in a Tris-d$_{11}$ D$_2$O buffer (Fig. 2 (b)), which only shows very low $^1$H signal intensities mainly due to non-deuterated Tris (chemical shift $\delta \approx$ 3.5 ppm) and traces of H$_2$O in D$_2$O (HDO, $\delta \approx$ 4.7 ppm).

Dioxane is an excellent viscosity control tracer in protein $^1$H-PFG-NMR, since it only shows one singlet peak, high solubility in H$_2$O and D$_2$O, is not charged, and does not interact with proteins.[24] To explore tracers of different sizes, we also studied 18-crown-6, which is a crown ether, a chemical analogue of dioxane, and shares all its advantages. The chemical shifts of dioxane and 18-crown-6 are similar, but spectrally well separated and therefore allow simultaneous determination of their diffusion coefficients (Fig. 2 (c) and (d)).

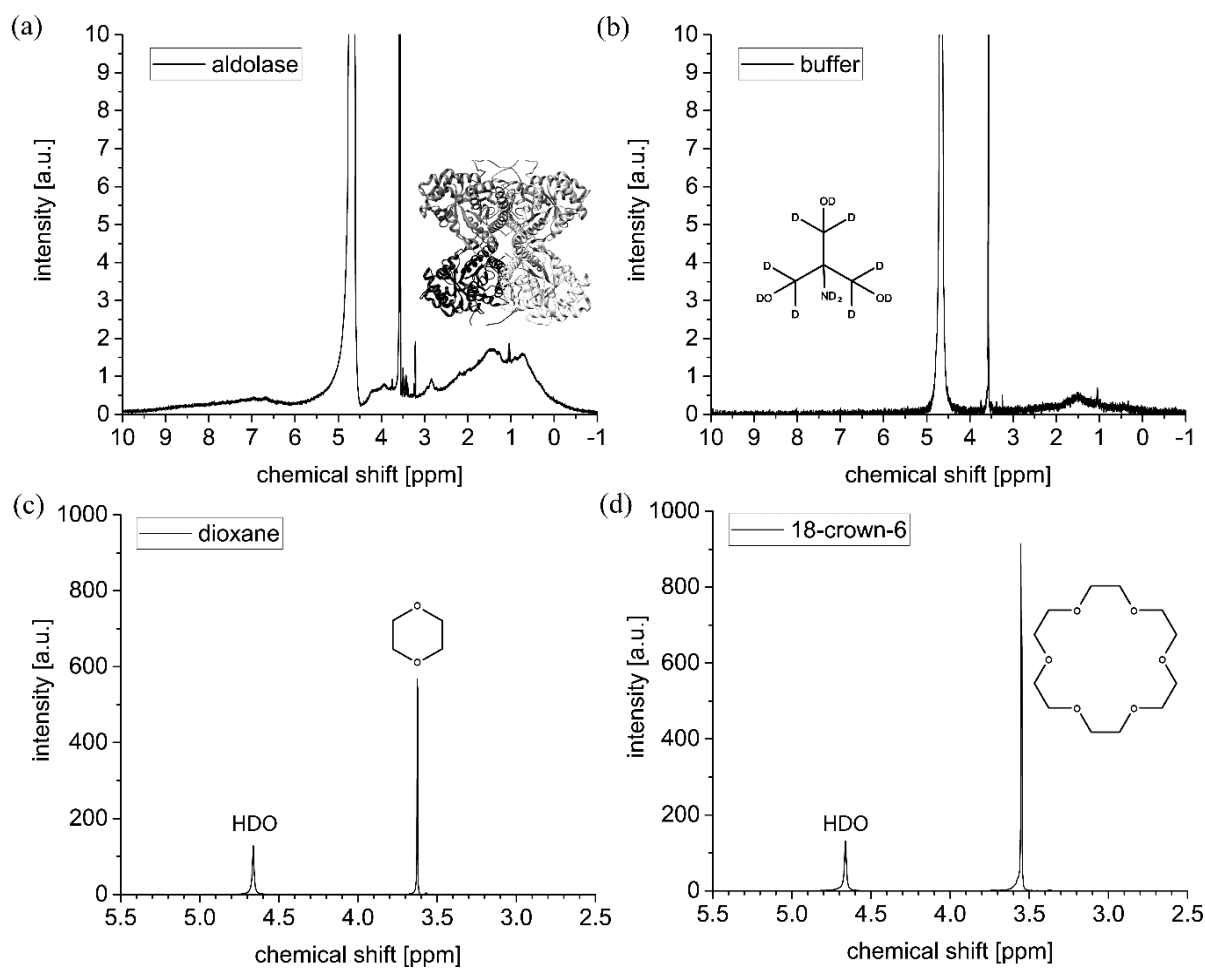

FIG. 2. $^1$H-NMR spectra and structure of the chemicals used in this study. (a) 11.9 µM ALD in Tris-$d_{11}$ buffer. (b) Tris-$d_{11}$ buffer alone. Tracer molecules dioxane (5 mg/mL) (c) and 18-crown-6 (5 mg/mL) (d) both containing residual HDO in Tris-$d_{11}$ buffer.

## II. Experimental Details

Materials

Fructose-bisphosphate aldolase (ALD) (product number A2714), triosephosphate isomerase type X (TPI) and α-glycerophosphate dehydrogenase type X (GDH) were all purchased from Sigma-Aldrich as lyophilized powders from rabbit muscle. The following chemicals D-fructose 1,6-bisphosphate (FBP) trisodium salt hydrate; anhydrous 1,4-dioxane; 18-crown-6; $D_2O$; 1 M tris(hydroxymethyl-d3)amino-d2-methane (Tris-$d_{11}$) solution in $D_2O$; 35 wt. % deuterium chloride in D$_2$O; and 30 wt. % sodium deuteroxide in D$_2$O were also purchased from Sigma-Aldrich. Anhydrous sodium pyrophosphate (PP) was purchased from Alfa Aesar.

Apart from ALD, the chemicals were used without further purification. ALD was purified as follows: ALD aggregates were removed by size-exclusion chromatography in Tris buffer using a HiPrep 16/60 Sephacryl S-300 HR column (GE Healthcare). The purified fraction of the ALD tetramer was then used in Tris-d$_{11}$ buffer for all subsequent NMR and activity experiments. Details on buffer preparation, chromatography and buffer exchange can be found in the supplementary information.

Enzyme activity assay

The activity of ALD was measured in the same solution (Tris-d$_{11}$ D$_2$O buffer) and at the same temperature (25 °C) as in the NMR experiments. A coupled activity assay of ALD with TPI and GDH has been performed to determine $K_M$ and $k_{cat}$ of the purified ALD. The activity of ALD was observed by monitoring the consumption of NADH through its 340 nm absorption band with a CARY 4000 UV/Vis spectrometer. The Michaelis-Menten analysis of the activity data as well as additional control measurements can be found in the supplementary information.

NMR and PFG-NMR experiments

1D-NMR spectra of molecules and the enzyme were recorded on a 400 MHz Jeol ECZ400S spectrometer in Tris-d$_{11}$ buffer at room temperature.

Diffusion coefficients of the enzyme ALD and of the small "tracer" molecules (HDO, dioxane, 18-crown-6) were measured by PFG-NMR with a Bruker Avance III 400 MHz spectrometer. Magnetic field gradients were generated using a diff60 diffusion probe and a Great60 gradient amplifier (Bruker Biospin) with the Bruker software TopSpin 3.2. All diffusivities were measured using the pulsed gradient stimulated-echo sequence with spoiler gradient pulses.[33] Gradient steps and operating parameters are listed in the supplementary information. In order to improve the signal-to-noise ratio, spectra were averaged up to 128 times. The diffusion probe was cooled with water to a constant temperature of 25 °C. Owing to the good thermal contact, the sample temperature was stable to 25.0 °C ± 0.3 °C throughout the measurement.

The diffusion constant was extracted from the spectra by integration of the NMR intensities $I$ and by applying a linear regression to the plot of $\ln(I(G))$ vs $b(G)$ with the help of a custom MATLAB code. Each experiment was repeated three times. Sample sizes were always 500 μL. The standard deviation of the data presented here was calculated via Gaussian error propagation

submitted to arXiv:1812.08748 [physics.chem-ph] 20 Dec 2018    6

from the standard deviation of the slope (*D*) of the linear regression. The results of the linear regressions of all individual experiments are listed in the supplementary information. If necessary, PFG-NMR spectra were averaged for clearer presentation.

## III. Results

Diffusion measurements of solutions in the presence of active aldolase

An active enzyme may experience enhanced diffusion due to a number of mechanisms. One possibility is that the catalytic activity causes a "mixing" of the solution. All solvent molecules and any small tracer molecules present in the solution should then also be stirred by flows induced by the enzymatic reactions. It has been reported that molecules and particles (*R*=0.57 nm to 1 μm) show enhanced diffusion, when active fructose-bisphosphate aldolase (ALD) is present even at low (10 nM) concentrations. These FCS-based studies found a diffusion enhancement of $D/D_0$=1.262 ± 0.063 for a tracer with *R*=0.57 nm at 1 mM FBP.[16] This cannot be explained with heating of the solution, since the catalyzed reaction is endothermic.

NMR is a powerful technique to test whether there are indeed any measurable flows due to the reaction, as the absolute diffusion coefficients of the solvent, the enzyme, and the tracer molecules can be determined. We have used three tracer molecules with *R*=0.11, 0.23 and 0.42 nm (based on Eq. 1 and results from Fig. 3 (c)) that exhibit strong single-peak NMR spectra (see Fig. 3(a)). The three tracer molecules were used to monitor the solution, when 10 nM active ALD is present, which catalyzes the cleavage of D-fructose 1,6-bisphosphate (FBP) to glyceraldehyde 3-phosphate (G3P) and dihydroxyacetone phosphate (DHAP):

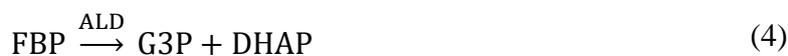

$$\text{FBP} \xrightarrow{\text{ALD}} \text{G3P} + \text{DHAP} \qquad (4)$$

The diffusion coefficients of all three molecules were monitored simultaneously and several substrate (FBP) concentrations were tested to probe whether the diffusion changes dependent on the enzyme's activity. The PFG-NMR spectrum (Fig. 3 (c)) shows the differences in attenuation of the signals of the different tracer molecules. The fastest diffusing molecule (HDO) exhibits strong attenuation even at low gradient field strengths visible at a chemical shift of $\delta \approx 4.7$ ppm. The peaks of dioxane and 18-crown-6 are spectrally very close to each other ($\delta \approx 3.7$ ppm and $\delta \approx 3.6$ ppm) due to their similar chemical makeup. Even though the signals lie close to each other, it was possible to analyze both independently, as they were spectrally resolved. Since fast repetition rates have been used, the mass concentration of dioxane was selected to be twice as much as the mass concentration of 18-crown-6 to compensate for differences in their spin-lattice relaxation. As expected, the attenuation of the stimulated echo of the smaller dioxane is stronger than for 18-crown-6.

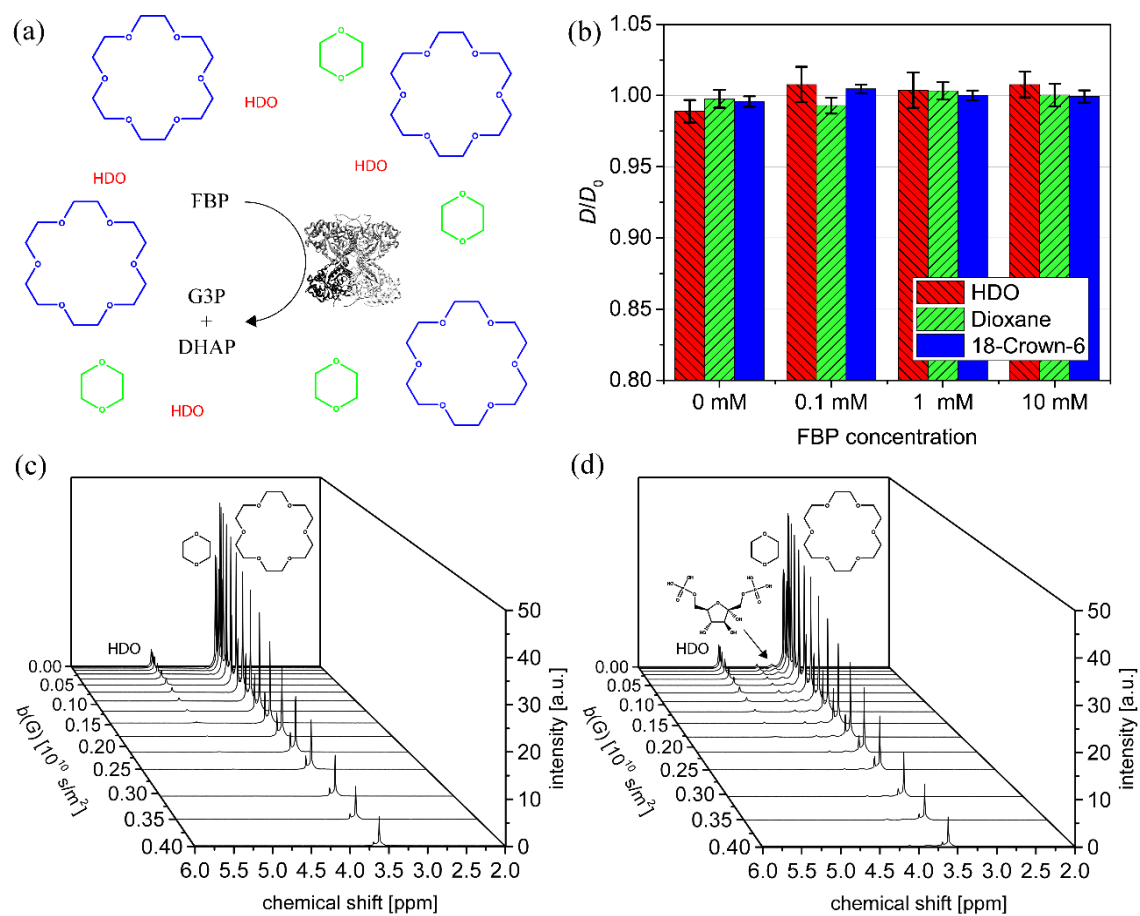

FIG. 3. Diffusion experiments of tracer molecules in active ALD solutions. (a) Schematic representation of reaction mixture containing 5 mg/mL dioxane (green), 2.5 mg/mL 18-crown-6 (blue), 10 nM ALD (grey) and different concentrations of FBP (not to scale). (b) Fractional diffusion coefficient change ($D/D_0$) of tracer molecules for different substrate (FBP) concentrations (tracer diffusion $D$ in the presence and $D_0$ in the absence of active ALD). Linear regressions in Fig. S3 and absolute diffusion coefficients in Tab. S3. Representative PFG-NMR spectra showing the signal decay of tracer molecules alone (c) and in an active ALD enzyme reaction mixture (d).

After the addition of ALD and FBP to the tracer mixture only small changes in the PFG-NMR spectrum (Fig. 3 (d)) are visible. Low signals of FBP can be seen around $\delta = 4$ ppm. ALD is not visible in the PFG-NMR spectrum due to its low concentration (10 nM). Since the addition of FBP has a small influence on the viscosity of the solution, we normalized the tracer diffusion coefficients $D$ of solutions containing 10 nM active ALD enzyme to those without enzyme ($D_0$), but containing the same FBP concentration. $D/D_0$ is presented for all experimental conditions in Fig. 3 (b). The linear regressions and diffusion coefficients of each individual experiment are listed in the supplementary information. We found no measurable diffusion enhancement. Even for HDO and 10mM FBP $D/D_0$ was $1.008 \pm 0.010$, where no diffusion change corresponds to $D/D_0=1$.

Diffusion of the enzyme aldolase during interaction with its inhibitor pyrophosphate

Remarkably, the endothermic enzyme ALD has also been reported to show enhanced diffusion when its inhibitor is present.[17] Similar effects of diffusion enhancement without substrate conversion have recently been reported for the enzymes urease[34] and hexokinase.[35] One possible explanation is that enzymes can exhibit significant structural changes upon interaction with inhibitors and hence show enhanced diffusion. We have recently shown that even though size-changes due to inhibitor interactions occur for the enzyme ATPase, several more dominant artefacts mask this effect in FCS.[11]

Enhanced diffusion has been reported for ALD in the presence of its inhibitor pyrophosphate (PP).[17] Illien et al. argue that the observed increase in diffusion of ALD ($D/D_0 \approx 1.2$) in the presence of its inhibitor PP is due to the repeated binding-unbinding fluctuations of the enzyme and the inhibitor. This is thought to cause structural changes that significantly affect the diffusion of the enzyme ALD.[28] Since equilibrium fluctuations are persistent, they are particularly amenable to NMR studies and permit high accuracy measurements. We have therefore performed PFG-NMR measurements over several minutes to probe the diffusion of ALD in the presence of PP, which does not possess a proton and can therefore be added in high concentration.

For this study, we focused on the aliphatic range of the ALD NMR spectrum with a chemical shift $\delta$ between 0.5 ppm and 1.6 ppm, since this region has the highest signal-to-noise ratio in the enzyme's NMR spectrum. In Fig. 4 (c) the PFG-NMR spectrum of 11.9 μM ALD is depicted. All ALD related signals between $\delta = 0$ and 5 ppm exhibit the same attenuation, which means that all these protons are diffusing with the same diffusion coefficient. Since very high gradients have been applied, the signals of the non-deuterated Tris and remaining HDO are only visible in the spectrum with the lowest gradient (4.242 T/m). From the PFG-NMR experiment of ALD alone, we were able to extract a diffusion coefficient $D_0$ for ALD of $(4.77 \pm 0.16)\ 10^{-11}\ m^2/s$.

The diffusion of ALD was then studied in the presence of 6 mM PP, which corresponds to a 500-fold excess of inhibitor per enzyme. No additional signals could be observed in the PFG-NMR spectrum. The attenuation of the ALD signal and the corresponding linear regressions are shown in Fig. 4 (b). The averaged $D/D_0$ of all experiments is $0.99 \pm 0.05$, where $D$ is the diffusion coefficient of ALD in the presence of PP and $D_0$ where no PP is present. Within the accuracy of the experiment we therefore find no increase or change in the diffusion coefficient of the enzyme ALD in the presence of its inhibitor PP.

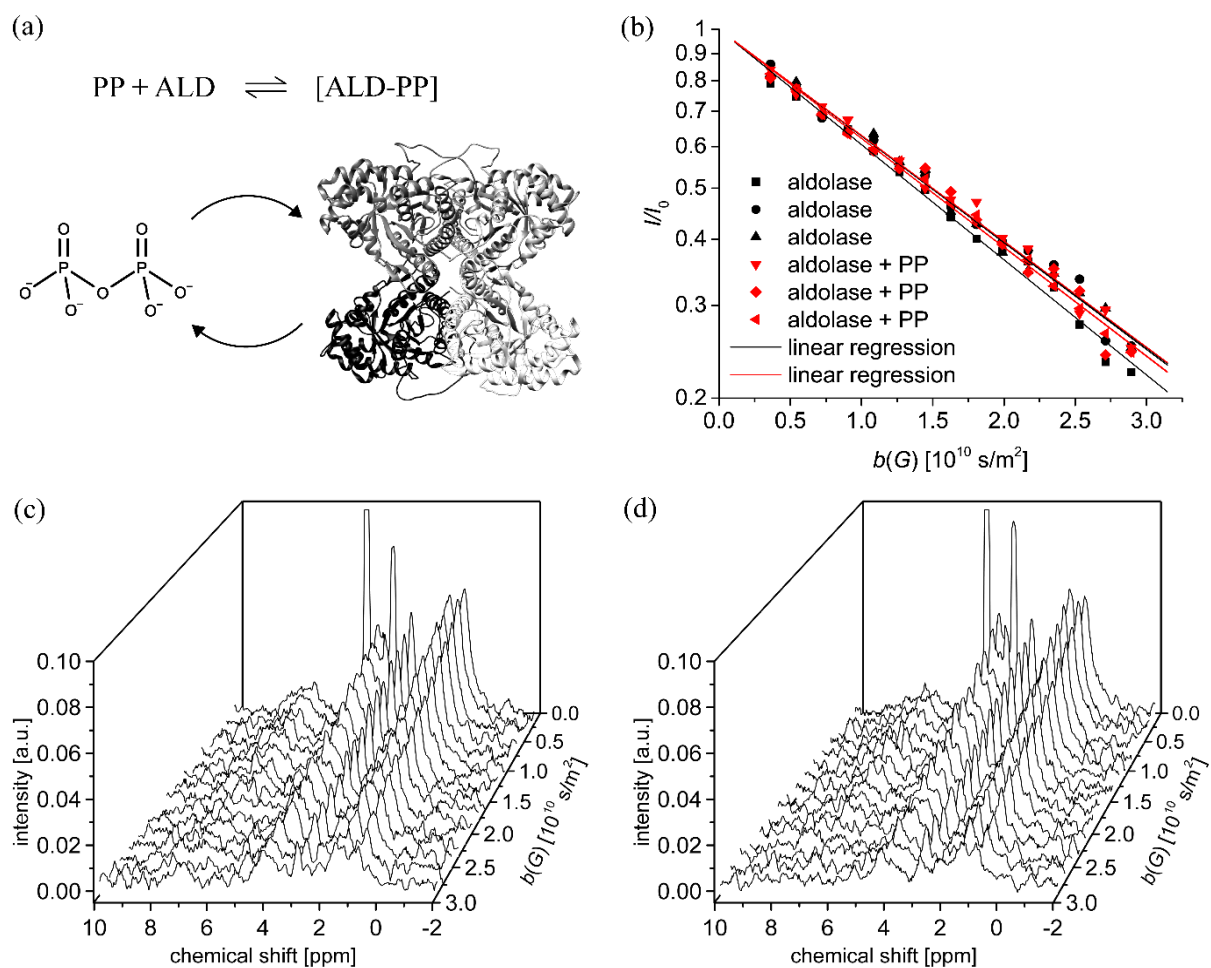

FIG. 4. Diffusion experiment of ALD interacting with its inhibitor PP. (a) Schematic of the chemical equilibrium in the solution containing 11.9 µM ALD and 6 mM PP in Tris-$d_{11}$ buffer (not to scale). (b) Linear regression of the decay of the aliphatic ALD signal between chemical shift δ=0.5 ppm and 1.6 ppm. The experiments were performed three times with and three times without the inhibitor PP. Representative PFG-NMR spectra showing the signal decay of ALD alone (c) and in the presence of its inhibitor PP (d).

Diffusion of the enzyme aldolase during substrate conversion

The strongest diffusion enhancements of the enzyme ALD have been reported during conversion of its substrate FBP. Fluorescence-based measurements report diffusion enhancements of $D/D_0 \approx 1.3$.[17,18] It is quite challenging to monitor an enzyme during substrate conversion with PFG-NMR, since 2D-NMR experiments are time consuming, because two parameters have to be sampled. Additionally, relatively high sample concentrations are needed especially for protein spectra. This requires large substrate concentrations, which must not deplete during the course of the measurement. We were able to obtain a satisfactory signal-to-noise ratio with a measuring time of only 4 min and 30 s, which we ran twice for each experiment. The turnover rate $k_{cat}$ of the purified ALD in Tris-$d_{11}$ buffer used in this study is 5.73 s$^{-1}$ ± 0.11 s$^{-1}$ (see supplementary information for details). The 50 mM FBP solution, which was used during our experiments, lasted for the full duration of the PFG-NMR experiment, i.e. 10.5 min (90 s of sample preparation/loading and 9 min for two measurements).

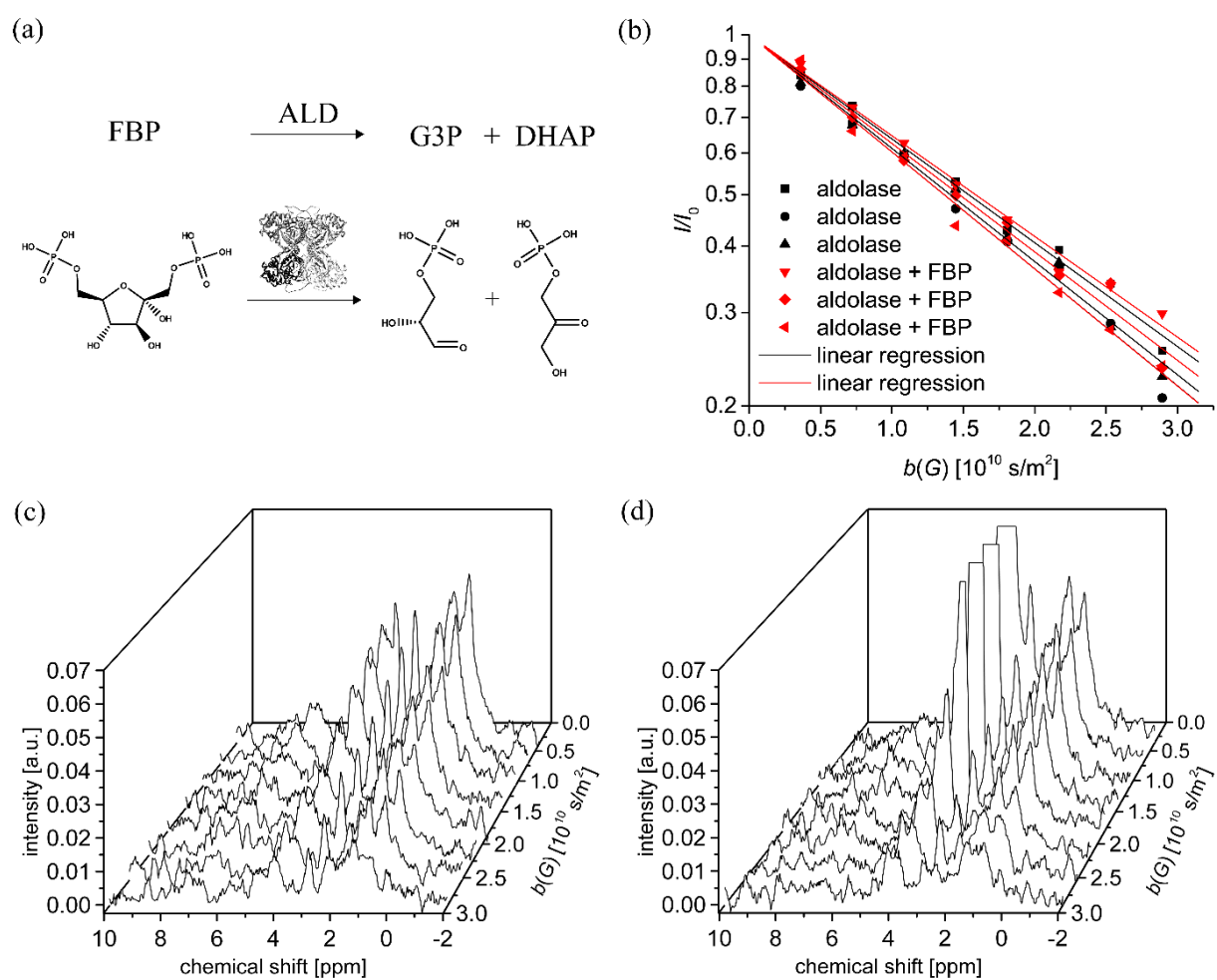

FIG. 5. Diffusion experiment of active ALD. (a) Schematic of the reaction solution containing 11.9 μM ALD and 50 mM FBP in Tris-$d_{11}$ buffer (not to scale). (b) Linear regression of the decay of the aliphatic ALD signal between chemical shift δ=0.5 ppm and 1.6 ppm. The experiments were performed three times with, and three times without the substrate FBP. Exemplary PFG-NMR spectra showing the signal decay of ALD alone (c) and during conversion of its substrate FBP (d).

The PFG-NMR spectrum of ALD recorded under the aforementioned conditions is shown in Fig. 5 (c). The measured diffusion coefficient $D_0$ is (4.8 ± 0.3) 10$^{-11}$ m$^2$/s as previously determined, but now with a higher standard deviation, which is expected since a smaller number of field gradients and fewer scans per gradient are used. If the substrate FBP is present in the solution, a very strong FBP signal around $\delta = 4$ ppm can be observed, which shows a faster attenuation than the ALD peaks. However, the analysis of the ALD diffusion is not affected by FBP, since there is no spectral overlap with the ALD signal in the range between δ=0.5-1.6 ppm (see Fig. S4). The ratio of the enzyme diffusion constant with substrate and of the enzyme alone $D/D_0$ is $0.98 \pm 0.09$. We do not find any diffusion enhancement in our PFG-NMR measurements for active aldolase.

## IV. Discussion

In all of our PFG-NMR measurements of the active enzyme aldolase we find no enhanced diffusion. We can only speculate what may have caused other, and in particular fluorescence correlation spectroscopy (FCS) measurements, to observe enhanced diffusion.[16,17,18] We recently listed a number of mechanisms that could give rise to apparent diffusion enhancement of enzymes in FCS measurements.[11] Of special interest here is the possible dissociation of ALD under substrate conversion.[36] The low concentrations used in FCS studies may exacerbate the problem of enzyme dissociation. In Fig. S1 (a) the chromatogram shows that many higher order aggregates are present in an ALD product from Sigma Aldrich. For this reason, we have used analytical size-exclusion chromatography to purify the ALD product prior to the measurements. The use of only molecular weight cutoff filters or short spin columns may therefore not be sufficient to exclude the presence of impurities like protein aggregates and unbound fluorophores, which may distort FCS measurements. A recent dynamic light scattering study, that also finds no diffusion enhancement of ALD during substrate conversion, uses a different, higher quality ALD product from Sigma-Aldrich,[37] which may support the hypothesis that aggregates have been present in previous FCS studies.

## V. Conclusion

In summary, we were able to show that PFG-NMR is a useful technique to measure absolute diffusion coefficients of active enzyme solutions. Even though only results in deuterated buffer have been presented in this study, working in protonated solvents is also possible with appropriate NMR solvent suppression sequences (e.g. WATERGATE).[38,32] Due to the high accuracy of the PFG-NMR method even small changes in the diffusion coefficient can be resolved. This is helpful, as it is currently debated whether enzymes self-propel when they are active, i.e. exhibit enhanced diffusion due to their activity. Recent fluorescence-based observations suggest enhancements of the diffusion coefficient of enzymes, when they turn over substrate and/or when an inhibitor is present. This would mean that enzymes are the smallest chemical motors. However, it has recently been shown[11] that the reported enhancements for ATPase[39] and phosphatase[15] are not due to activity, but are due to artefacts in the fluorescence measurements.

Another enzyme, for which enhanced diffusion due to activity has been reported using fluorescence correlation spectroscopy (FCS), is fructose-bisphosphate aldolase (ALD). ALD is a particularly interesting enzyme, because the reaction it catalyzes is endothermic, and because even interactions with its inhibitor are reported to cause increases in diffusion.[16,17,18] During the preparation of this manuscript a study utilizing dynamic light scattering has been published, where the authors find no diffusion enhancement of aldolase during conversion with its substrate.[37] Our PFG-NMR study is in agreement with this latest result – we do not detect any change or increase in the diffusion of aldolase when ALD converts its substrate (FBP). In addition, we find no measurable increase in the diffusion of the solution itself, when aldolase is active and turns over substrate. We also do not find any change in the diffusion of aldolase in the presence of its inhibitor (PP). Our PFG-NMR measurements therefore summarily find no enhanced diffusion when the enzyme aldolase is catalytically active. It is interesting to ask what the smallest entity (molecule, enzyme or nanoparticle) is that shows enhanced diffusion and that can act as a self-propelled chemical motor. Further experiments are necessary to address this question.

## Supplementary Information

See supplementary information for details on enzyme purification, activity assay and complete results of the PFG-NMR experiments as well as the 1D-$^1$H-NMR of FBP.

## Acknowledgments

The Authors thank Christine Mollenhauer for her support and help with protein purification. This work was supported by the DFG in the priority program SPP 1726 as project 253407113 (P.F.).

# Supplementary Information for "Absolute diffusion measurements of active enzyme solutions by NMR"

Jan-Philipp Günther,[1,2] Günter Majer,[1,*] Peer Fischer[1,2,*]

[1]Max Planck Institute for Intelligent Systems, Heisenbergstr. 3, 70569 Stuttgart, Germany

[2]Institute of Physical Chemistry, University of Stuttgart, Pfaffenwaldring 55, 70569 Stuttgart, Germany

*Authors to whom correspondence should be addressed: Günter Majer (majer@is.mpg.de), Peer Fischer (fischer@is.mpg.de)

## Aldolase purification

Fructose-bisphosphate aldolase (ALD) was dissolved in Tris buffer (150 mM Tris, 100 mM NaCl, pH 7.4, in $H_2O$) at a concentration of 5 mg/mL or higher. Up to 5 mL of this solution were injected onto a HiPrep 16/60 Sephacryl S-300 HR column (GE Healthcare), which previously has been flushed with Tris buffer. The flow speed during the purification was 0.5 mL/min and the ALD tetramer was collected typically in the elution volume between 64 and 72 mL (Fig. S1 (a)). Purified ALD was then concentrated with Amicon Ultra-4 10k (Merck Millipore) molecular weight cutoff filters and the buffer was changed to a Tris-$d_{11}$ buffer with Micro Bio-Spin 6 (Bio-Rad) spin columns. The Tris-$d_{11}$ buffer contained 100 mM Tris-$d_{11}$ in $D_2O$, which was adjusted with a pH electrode and DCl and KOD to a pD of 7.4.[1] The concentration of the ALD solution in Tris-$d_{11}$ buffer was then determined via UV spectroscopy (Fig. S1 (b)) using the theoretical extinction coefficient of the ALD tetramer at 280 nm of $\varepsilon = 134,100\ (M\cdot cm)^{-1}$ (calculated[2] from the PDB entry 1zah[3]).

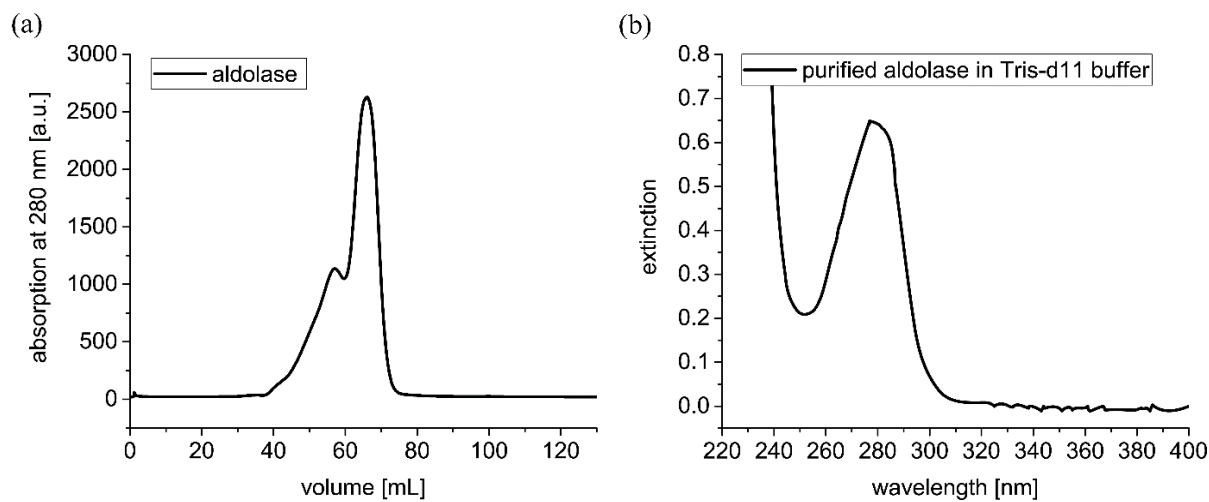

FIG. S1. Purification of ALD. (a) Chromatogram of ALD on a Sephacryl size exclusion column in Tris buffer. (b) UV extinction spectrum of aldolase tetramer fraction (64-72 mL) after buffer exchange to a Tris-$d_{11}$ buffer.

**Aldolase activity assay**

The ALD activity was determined in a coupled cascade assay of ALD, triosephosphate isomerase (TPI) and α-glycerophosphate dehydrogenase (GDH) by monitoring the consumption of NADH at 340 nm.[4] D-Fructose 1,6-bisphosphate (FBP) is cleaved by ALD to glyceraldehyde 3-phosphate (G3P) and dihydroxyacetone phosphate (DHAP) (see Eq. (S1)). TPI then converts G3P through isomerization into additional DHAP (see Eq. (S2)). The final reaction of the cascade is the reduction of DHAP to glycerophosphate, which is catalyzed by GDH (see Eq. (S3)). This consumes NADH, which leads to a diminishing UV extinction signal at 340 nm. The activity of ALD can be calculated from the slope of the UV signal of NADH over time. Two NADH are consumed per cleaved FBP.

Typically 33-200 nM ALD, 1.3 nM TPI, 59 nM GDH and 233 μM NADH were mixed with concentrations of FBP between 0 and 50 mM. The whole assay was performed in Tris-$d_{11}$ buffer to be comparable to the NMR experiments. The temperature during the activity assay was 24-25 °C. Three measurements were performed for each condition tested. A Michaelis-Menten analysis (Fig. S2) of the data was performed. This lead to a $k_{cat}$ of 5.73 s$^{-1}$ ± 0.11 s$^{-1}$ and a $K_M$ of 47.0 μM ± 8.8 μM. This is more than four times higher than those reported by Zhao et al.[5] As an additional control, the activity of ALD after all NMR experiments and prolonged exposure to room temperature was measured at 10 mM FBP, which yielded an activity of at least 2.19 s$^{-1}$ ± 0.09 s$^{-1}$ or higher. A similar measurement was performed for ALD after prolonged exposure to room temperature using 10 mM FBP in the presence of tracer molecule mixture (5 mg/mL dioxane and 2.5 mg/mL 18-crown-6), which lead to an activity of 2.01 s$^{-1}$ ± 0.08 s$^{-1}$.

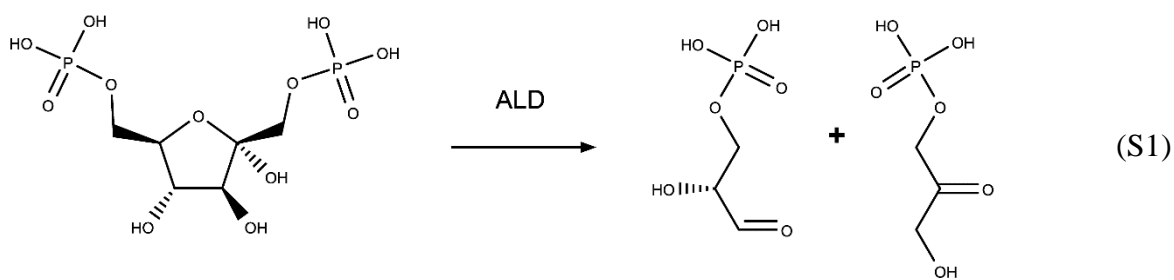 (S1)

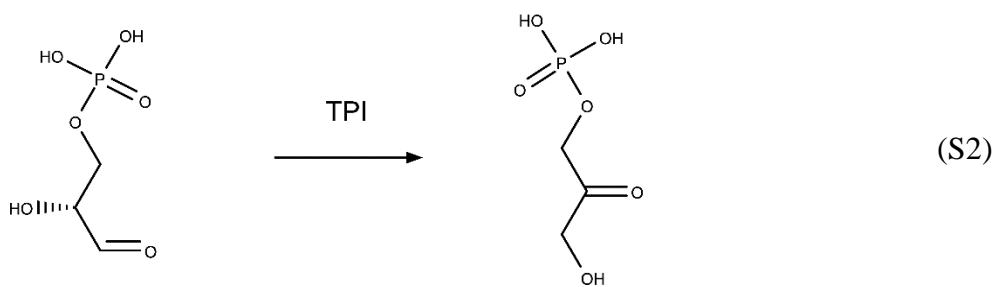 (S2)

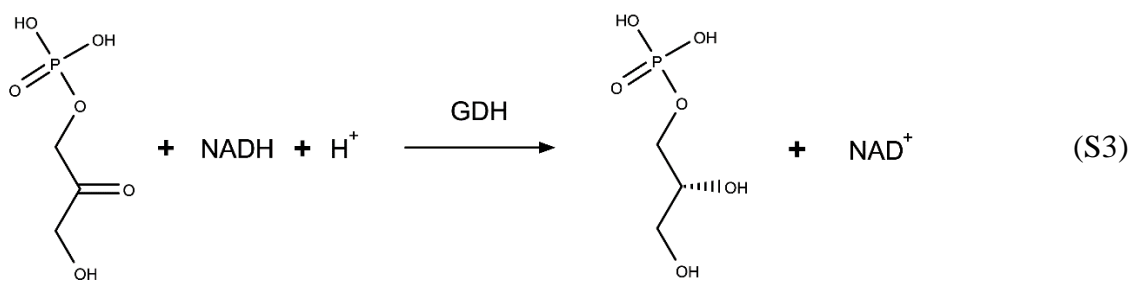 + NADH + H$^+$ → + NAD$^+$ (S3)

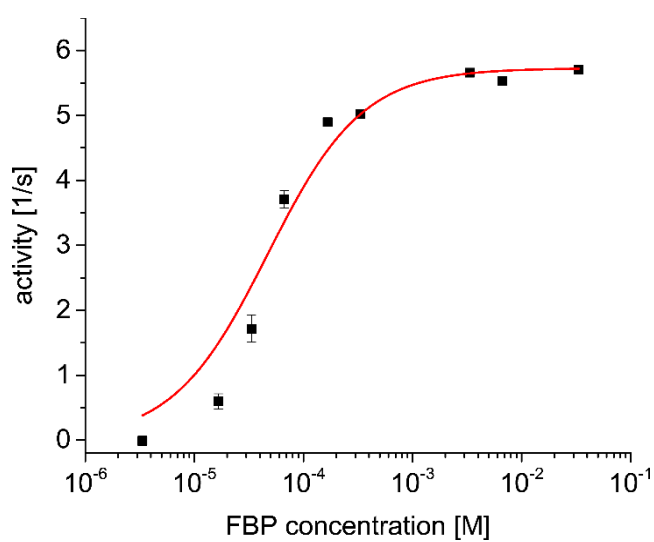

FIG. S2. ALD activity assay. Michaelis-Menten plot of ALD activity in Tris-d$_{11}$ buffer. Each data point results from three measurements.

# Detailed settings for the PFG-NMR experiments

For all PFG-NMR experiments a stimulated echo sequence has been used with the parameters listed in Tab. S1. Special care was taken to utilize a very fast pulse sequence for the experiments where the ALD diffusion was monitored during FBP turnover. All the gradients used in the experiments of this paper are listed in Tab. S2.

TAB. S1. Parameters for the PFG-NMR experiments presented in Fig. 3, 4 and 5.

| parameter | experiments | | |
| --- | --- | --- | --- |
| | tracers | ALD + PP | ALD + FBP |
| $\delta_G$ [ms] | 1 | 0.6 | 0.6 |
| $\Delta$ [ms] | 10 | 8 | 8 |
| length of FID [s] | 1.049 | 0.262 | 0.262 |
| recovery delay [s] | 1.00 | 0.50 | 0.25 |
| number of scans | 16 | 128 | 2x64 |
| number of dummy scans | 1 | 1 | 1 |
| number of gradients | 16 | 16 | 8 |
| total time | 8 min 50 s | 26 min 30 s | 9 min |

TAB. S2. Gradients $G$ used in the PFG-NMR experiments presented in Fig. 3, 4 and 5.

| gradient number | experiments | | |
| --- | --- | --- | --- |
| | tracers $G$ [T/m] | ALD + PP $G$ [T/m] | ALD + FBP $G$ [T/m] |
| 1 | 0.100 | 3.000 | 4.242 |
| 2 | 0.250 | 4.242 | 5.999 |
| 3 | 0.400 | 5.195 | 7.347 |
| 4 | 0.550 | 5.999 | 8.484 |
| 5 | 0.700 | 6.707 | 9.485 |
| 6 | 0.850 | 7.347 | 10.390 |
| 7 | 1.000 | 7.936 | 11.220 |
| 8 | 1.150 | 8.484 | 12.000 |
| 9 | 1.300 | 8.998 | |
| 10 | 1.450 | 9.485 | |
| 11 | 1.600 | 9.948 | |
| 12 | 1.750 | 10.390 | |
| 13 | 1.900 | 10.810 | |
| 14 | 2.100 | 11.220 | |
| 15 | 2.250 | 11.620 | |
| 16 | 2.400 | 12.000 | |

## Detailed results of the PFG-NMR experiments

For all PFG-NMR experiments the signals of the individual molecules were integrated for each gradient to calculate $I(G)$. The intervals and gradients used for those integrations are listed at the bottom of the individual tables. Subsequently, a linear regressions of $\ln(I(G))$ vs $b(G)$ was performed, which resulted in the values of the intercept $\ln(I_0)$ and the slope $D$. All the results for $D$ are listed together with the standard deviation and the $R^2$ value obtained from the linear regression. All linear regression had an $R^2 > 0.97$.

<u>Diffusion measurements of solutions in the presence of active aldolase</u>

Fig. S3 presents the data and linear regression of all tracer experiments performed. The differences in the signal attenuation for each molecule are seen in the figures. The results of the linear regression are listed in Tab. S3. Here all $R^2$ values are greater then 0.998.

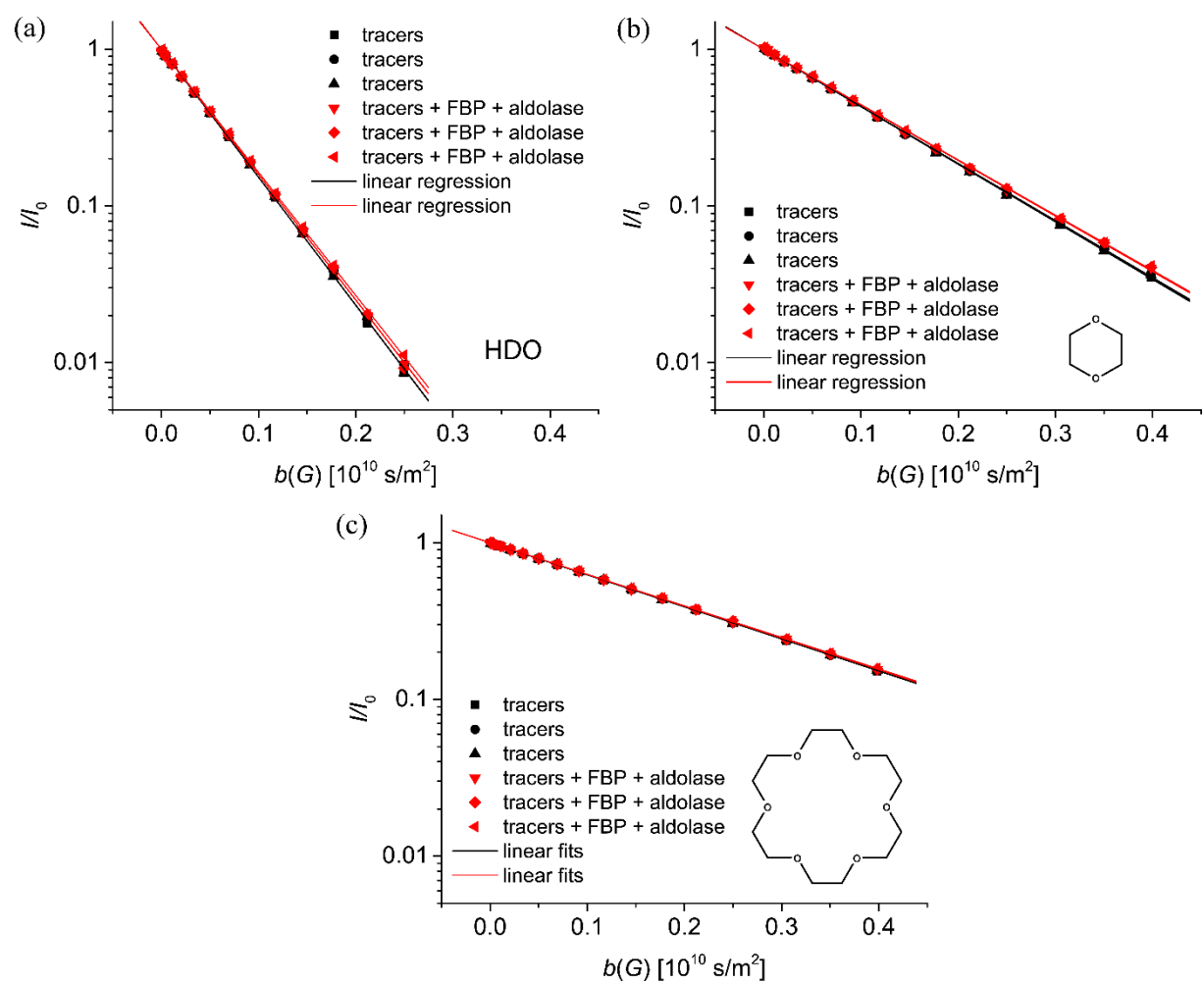

FIG. S3. Linear regressions of the tracer molecule decay during the experiments presented in Fig. 3. (a) HDO signal between δ=4.680 ppm and 4.750 ppm. (b) dioxane signal between δ=3.680 ppm and 3.705 ppm. (c) 18-crown-6 signal between δ=3.604 ppm and 3.635 ppm. 5 mg/mL dioxane and 2.5 mg/mL 18-crown-6 were used for the control experiments (black). Additionally, 10 mM FBP and 10 nM ALD were present during the active ALD experiments (red).

TAB. S3. Diffusion coefficients of tracer molecules from PFG-NMR experiments performed in active ALD solutions.

| $c$(FBP) [mM] | $c$(aldolase) [nM] | $D$(HDO)* [$10^{-10}$ m$^2$/s] | $R^2$ | $D(C_4H_8O_2)$** [$10^{-10}$ m$^2$/s] | $R^2$ | $D(C_{12}H_{24}O_6)$*** [$10^{-10}$ m$^2$/s] | $R^2$ |
|---|---|---|---|---|---|---|---|
| 0 | 0 | 18.81 ± 0.13 | 0.9996 | 8.419 ± 0.025 | 0.9999 | 4.714 ± 0.014 | 0.9999 |
| 0 | 0 | 18.42 ± 0.06 | 0.9999 | 8.380 ± 0.042 | 0.9997 | 4.697 ± 0.012 | 0.9999 |
| 0 | 0 | 18.78 ± 0.12 | 0.9996 | 8.451 ± 0.043 | 0.9997 | 4.711 ± 0.016 | 0.9999 |
| 0 | 10 | 18.43 ± 0.08 | 0.9998 | 8.391 ± 0.042 | 0.9997 | 4.697 ± 0.009 | 1.0000 |
| 0 | 10 | 18.39 ± 0.08 | 0.9999 | 8.409 ± 0.025 | 0.9999 | 4.685 ± 0.011 | 0.9999 |
| 0 | 10 | 18.57 ± 0.13 | 0.9995 | 8.391 ± 0.042 | 0.9997 | 4.682 ± 0.011 | 0.9999 |
| 0.1 | 0 | 19.10 ± 0.22 | 0.9988 | 8.399 ± 0.035 | 0.9998 | 4.714 ± 0.010 | 0.9999 |
| 0.1 | 0 | 17.95 ± 0.12 | 0.9996 | 8.295 ± 0.032 | 0.9998 | 4.624 ± 0.008 | 1.0000 |
| 0.1 | 0 | 18.26 ± 0.11 | 0.9997 | 8.494 ± 0.019 | 0.9999 | 4.678 ± 0.014 | 0.9999 |
| 0.1 | 10 | 18.62 ± 0.15 | 0.9994 | 8.341 ± 0.044 | 0.9997 | 4.699 ± 0.009 | 1.0000 |
| 0.1 | 10 | 18.16 ± 0.07 | 0.9999 | 8.284 ± 0.037 | 0.9998 | 4.669 ± 0.005 | 1.0000 |
| 0.1 | 10 | 18.95 ± 0.23 | 0.9987 | 8.387 ± 0.028 | 0.9999 | 4.713 ± 0.011 | 0.9999 |
| 1 | 0 | 18.48 ± 0.18 | 0.9992 | 8.301 ± 0.037 | 0.9998 | 4.669 ± 0.014 | 0.9999 |
| 1 | 0 | 18.60 ± 0.09 | 0.9998 | 8.359 ± 0.030 | 0.9998 | 4.679 ± 0.015 | 0.9999 |
| 1 | 0 | 18.48 ± 0.07 | 0.9999 | 8.306 ± 0.036 | 0.9998 | 4.700 ± 0.011 | 0.9999 |
| 1 | 10 | 18.51 ± 0.14 | 0.9995 | 8.366 ± 0.033 | 0.9998 | 4.696 ± 0.010 | 0.9999 |
| 1 | 10 | 18.66 ± 0.27 | 0.9981 | 8.307 ± 0.040 | 0.9997 | 4.641 ± 0.009 | 1.0000 |
| 1 | 10 | 18.61 ± 0.15 | 0.9995 | 8.377 ± 0.035 | 0.9998 | 4.712 ± 0.006 | 1.0000 |
| 10 | 0 | 18.57 ± 0.19 | 0.9991 | 8.140 ± 0.042 | 0.9997 | 4.649 ± 0.016 | 0.9999 |
| 10 | 0 | 17.76 ± 0.05 | 0.9999 | 8.007 ± 0.043 | 0.9997 | 4.559 ± 0.006 | 1.0000 |
| 10 | 0 | 18.16 ± 0.07 | 0.9999 | 8.283 ± 0.041 | 0.9997 | 4.745 ± 0.015 | 0.9999 |
| 10 | 10 | 18.40 ± 0.09 | 0.9998 | 8.162 ± 0.052 | 0.9995 | 4.663 ± 0.016 | 0.9999 |
| 10 | 10 | 18.42 ± 0.14 | 0.9995 | 8.155 ± 0.046 | 0.9996 | 4.647 ± 0.011 | 0.9999 |
| 10 | 10 | 18.08 ± 0.10 | 0.9997 | 8.120 ± 0.048 | 0.9996 | 4.633 ± 0.018 | 0.9998 |

*gradients 1-13, δ=4.680-4.750 ppm
**gradients 1-16, δ=3.680-3.705 ppm
***gradients 1-16, δ=3.604-3.635 ppm

## Diffusion of the enzyme aldolase during interaction with its inhibitor pyrophosphate

Diffusion coefficients are listed in Tab. S4 for all experiments involving the interaction of ALD with its inhibitor PP.

TAB. S4. Diffusion coefficients of ALD PFG-NMR experiments performed during interaction with its inhibitor PP.

| $c$(PP) [mM] | $c$(aldolase) [μM] | $D$(aldolase)* [$10^{-11}$ m$^2$/s] | $R^2$ |
|---|---|---|---|
| 0 | 11.9 | 5.03 ± 0.16 | 0.989 |
| 0 | 11.9 | 4.63 ± 0.18 | 0.985 |
| 0 | 11.9 | 4.65 ± 0.14 | 0.990 |
| 6 | 11.9 | 4.62 ± 0.15 | 0.990 |
| 6 | 11.9 | 4.75 ± 0.21 | 0.980 |
| 6 | 11.9 | 4.76 ± 0.11 | 0.995 |

*gradients 2-16, δ=0.5-1.6 ppm

## Diffusion of the enzyme aldolase during substrate conversion

In Fig. S4 the 1D-$^1$H-NMR spectrum of FBP, the substrate of ALD, is shown. FBP shows no proton NMR signal in the interval between δ=0.5 ppm and 1.6 ppm, and therefore does not interfere with the aliphatic ALD signal analyzed in this paper. It is also seen that the FBP spectrum in Fig. S4 exhibits more peaks than would be expected for a simple furanose structure. The additional peaks arise because FBP is in equilibrium between its furanose and keto form. Interestingly, the keto form of FBP shows a higher reaction rate with ALD than its furanose form.[6]

The results of the linear regression of all ALD diffusion experiments during substrate conversion are listed in Tab. S5.

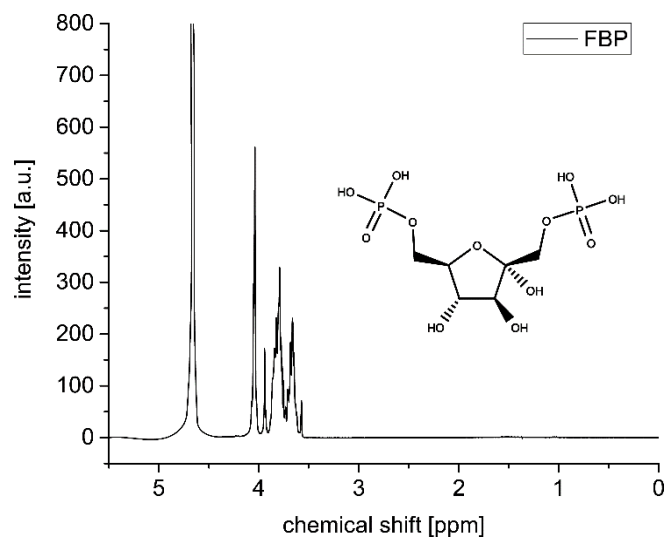

FIG S4. $^1$H-NMR spectrum of 50 mM FBP in Tris-d$_{11}$ buffer and chemical structure of FBP.

TAB. S5. Diffusion coefficients of ALD PFG-NMR experiments performed during conversion of its substrate FBP.

| $c$(FBP) [mM] | $c$(aldolase) [µM] | $D$(aldolase)* [$10^{-11}$ m$^2$/s] | $R^2$ |
|---|---|---|---|
| 0 | 11.9 | 4.50 ± 0.16 | 0.989 |
| 0 | 11.9 | 5.07 ± 0.18 | 0.985 |
| 0 | 11.9 | 4.91 ± 0.14 | 0.990 |
| 50 | 11.9 | 4.37 ± 0.15 | 0.990 |
| 50 | 11.9 | 4.70 ± 0.21 | 0.980 |
| 50 | 11.9 | 5.07 ± 0.11 | 0.995 |

*gradients 1-8, δ=0.5-1.6 ppm